\documentclass[useAMS,usenatbib]{mn2e}

\newcommand{\simgt}{\mathrel{\raise0.35ex\hbox{$\scriptstyle >$}\kern-0.6em
\lower0.40ex\hbox{{$\scriptstyle \sim$}}}}
\newcommand{\simlt}{\mathrel{\raise0.35ex\hbox{$\scriptstyle <$}\kern-0.6em
\lower0.40ex\hbox{{$\scriptstyle \sim$}}}}

\input epsf

\begin{document}

\title[Accretion flows in early--type galaxies
 and CMB experiments]{ Radio Emission from Early-type Galaxies
and CMB Experiments}

\author[E.~Pierpaoli and R.~Perna] 
{Elena Pierpaoli${}^{1,2}$ and Rosalba Perna${}^{2}$\\
${}^1$Physics Department,
 Princeton University, Princeton, NJ, 08544~~USA\\
${}^2$ Astronomy Department,
 Princeton University, Princeton, NJ, 08544~~USA\\
}

\date{Accepted ... ;
      Received ... ;
      in original form ...}

\pagerange{000--000}

\maketitle

\begin{abstract}

We investigate the possible contribution from the emission of
accretion flows around supermassive black holes in early type galaxies
to current measurements of the Cosmic Microwave Background (CMB) at
radio frequencies.  We consider a range of  luminosities
suggested by targeted radio observations and accretion models and
compute the residual contribution of these sources
to the spectrum and bispectrum of the
observed CMB maps.  As for high--resolution CMB experiments, we find
that the unresolved component of these sources could make up to $\sim
40-50\%$ of the observed CBI and BIMA power spectrum at $l > 2000$. As
a consequence, the inferred $\sigma_8^{SZ}$ value could be biased high
by up to $6-7\%$.  As for all sky experiments, we find that the
contribution of accretion-flow sources to the WMAP bispectrum is at
the 2-3 per cent level at most.  At the flux limit that Planck will
achieve, however, these sources may contribute up to 15 per cent of
the bispectrum in the 60--100 GHz frequency range.  Moreover, Planck
should detect hundreds of these sources in the 30--300 GHz frequency
window.  These detections, possibly coupled with galaxy type
confirmation from optical surveys, will allow number counts to put
tighter constraints on early-type galaxies radio luminosity and
accretion flows properties.  These sources may also contribute up to
the 30 per cent level to the residual radio sources power spectrum in
future high--resolution SZ surveys (like ACT or APEX) reaching mJy
flux limits.

\end{abstract}

\begin{keywords}
accretion -- cosmology: cosmic microwave background --
galaxies: elliptical -- galaxies:nuclei 
 --  infrared:galaxies
\end{keywords}

\section{Introduction}

Cosmic microwave background (CMB) experiments have been able to
measure cosmological parameters to an unprecedented level of accuracy
\nocite{Spergel03}({Spergel} {et~al.} 2003).  For this to be possible,
however, the contribution to the primary signals must be disentangled
from other astrophysical emissions at the observed frequencies.
Therefore, foreground identification and removal is fundamental.

One of the major sources of contamination at small scales is
constituted by point sources, and several studies have been carried
out to estimate their contribution to CMB anisotropy experiments
\nocite{Toffo98,WM03,Argueso03}({Toffolatti} {et~al.} 1998; {White} \&
{Majumdar} 2004; {Arg{\" u}eso}, {Gonz{\' a}lez-Nuevo} \& {Toffolatti}
2003).  Generally, source counts are determined in the radio band,
through deep VLA surveys down to $\mu$Jy levels at 1.41, 4.86 and 8.44
GHz.  These counts are then extrapolated to the higher range of
frequencies relevant for CMB experiments
\nocite{Toffo98,tof99,dezotti,gaw}({Toffolatti} {et~al.} 1998;
{Toffolatti} \& {et al.} 1999; {De Zotti} {et~al.} 2000; {Gawiser} \&
{Smoot} 1997).  This extrapolation provides a reasonable estimate of
the contribution from the ``steep'' and ``flat'' spectrum sources
(with $F_\nu\propto \nu^{-\alpha}$, and $\alpha\ge 0$, such as compact
radio galaxies and radio loud quasars), but it strongly
under-represents an important contribution from a class of sources
with inverted spectra ($\alpha < 0$; e.g. De Zotti et al.  2000).

Inverted-spectrum sources, such as GHz peaked sources (GPS;
\nocite{odea,guerra}{O'Dea} (1998); {Guerra}, {Haarsma} \& {Partridge}
(1998)),  bright (flux $\sim 1-10$ Jy) and rare, are generally associated
with bright active galaxies or quasars at high redshifts.  Their
contribution to the CMB experiments was studied by \nocite{dezotti}{De
Zotti} {et~al.} (2000).  These sources, being bright and rare, are
easily identified in the CMB maps, and removed.

There is however another class of inverted-spectrum sources, much
fainter (flux $\sim 1$ mJy) but much more common, associated with
emission from the nuclei of nearby galaxies.  Radio continuum surveys
(at $\nu < $ 8 GHz) of elliptical and S0 galaxies have shown that
the sources in radio--quiet galaxies tend to be extended but with a
compact component with relatively flat or slowly rising radio spectra.
 Recent VLA studies at
high radio frequencies (up to 43 GHz), although carried out only on a
limited sample of objects, have shown that all of the observed compact
cores have spectra rising up to $\sim 20-30$ GHz.  These sources of
radio emission are believed to result from the process of accretion of
gas into the supermassive black holes likely ubiquitous in
the centers of galaxies \nocite{magorrian}({Magorrian} {et~al.} 1998).

A popular model to describe the broadband spectral energy
distribution from the accretion flows around these supermassive black
holes is the advection-dominated accretion flow (ADAF) model
\nocite{rees,ny}({Rees} {et~al.} 1982; {Narayan} \& {Yi} 1994).
Within the context of this model, the foreground contribution to the
CMB experiments, and in particular to the {\em Planck Surveyor}, was
studied by \nocite{Perna00}{Perna} \& {Di Matteo} (2000).  In
particular, they estimated the contribution of these sources both
within the ``standard'' ADAF model, where the accretion rate is
independent of the radius $R$ of the flow (implying that all the mass
is accreted into the black hole (BH)), and within the context of the
ADAF model with winds, where not all the mass is accreted by the BH,
but some is lost into winds \nocite{bb,ny95}({Blandford} \& {Begelman}
1999; {Narayan} \& {Yi} 1995a). In this case the radio emission is suppressed because 
winds remove mass from the inner regions of the flow, where the synchrotron
radiation giving rise to the radio emission is produced.
Detailed modeling has only been possible for a handful of sources so far,
as it requires multi wavelength, high-resolution data. 
Observations have yielded somewhat mixed evidence: in some cases
the emission is consistent with ADAFs with winds (i.e.
suppressed radio emission; Di Matteo et al. 2000), while in others it is higher than
the standard model with no wind \nocite{dm01}(Di Matteo et al. 2001), probably because of extra
power output by jets associated with the accretion flow. 

While studies of this class of sources of radio emission so far have
been limited by low number statistics, \nocite{Perna00}{Perna} \& {Di
Matteo} (2000) showed that CMB experiments with the future {\em Planck
surveyor} mission can provide interesting constraints
from a statistical point of view.  At the same time, these sources can be an
important contaminant to CMB maps.

In this paper, we compute the contribution of the emission from
accretion flows in early-type galaxies to the current signal of CBI,
BIMA, WMAP.  We show that the signal from these unresolved
sources is able to influence current limits on the power spectrum
normalization $\sigma_8$.  Finally, we make predictions on the power
spectrum from residual sources in future high--resolution CMB
experiments and on the bispectrum in all--sky experiments like Planck
and WMAP.

\section{Microwave emission from accretion flows in elliptical galaxies} 

As discussed above, there is evidence for the existence of supermassive massive
black holes (BHs) at the center of galaxies. Inferred black hole masses 
appear to be proportional to the mass of the bulge component of
their host galaxies. Therefore, central BH masses are expected to be
much larger in elliptical than in spiral galaxies. 
Independently of the details of the model for the production of the
radio/microwave emission, this radiation is expected to be some fraction of the
accretion energy, which is proportional to the mass of the black hole.
Therefore the contribution from the nuclei of the ellipticals is 
expected to dominate that from the nuclei of the spirals, which we
neglect here. Similarly, among the population of ellipticals, the main contribution 
derives from  the most massive galaxies. To make a conservative
estimate of the emission from these galaxies, we consider only the
bright end of the distribution, with $L \simgt L_*$. Studying a sample of
nearly 9000 ellipticals in the Sloan Digital Sky Survey, \nocite{ber}{Bernardi} {et~al.} (2003) 
estimated a comoving number density $\Phi_*=(5.8\pm 0.3)\times 10^{-3} h^3$
Mpc$^{-3}$. The redshift evolution of their sample appeared consistent
with the law $\Phi_*(z)=10^{0.4Pz}\Phi_*(0)$ ($P\approx -2$), found by
\nocite{lin}{Lin} {et~al.} (1999) in their sample of galaxies drawn from the Canadian Network
for Observational Cosmology Field Galaxy Redshift Survey. 

For a cosmological population of sources with intrinsic luminosity function $f(L_\nu)$
and redshift evolution $\Phi_*(z)$, the differential
number counts are given by
\begin{eqnarray}
& &\frac{dn(S_\nu)}{dS_\nu} = \nonumber \\
& & = \int_0^\infty dL_\nu \left[f(L_\nu) \Phi(z)\;
\frac{dV(z)}{dz} \left|\frac{dz}{dS_\nu}(z,L_\nu)\right|\right]_{z(S_\nu,L_\nu)}
\label{eq:dnds}
\end{eqnarray}
where  $dV(z)/dz$ is the comoving volume. 
In a flat cosmology with a cosmological constant it is given by
\begin{equation}
\frac{dV(z)}{dz} = 4\pi {D^2(z)}\frac{dD(z)}{dz}\;
\end{equation}
where $D(z)$ is the comoving distance, 
\begin{equation}
D(z) = {c \over H_0} \int_0^z {dz' \over \sqrt{(1+\Omega_mz')(1+z')^2-\Omega_\Lambda(2z'+z'^2)} } \;,
\end{equation}
We assume a cosmological model with $\Omega_m=0.3$, $\Omega_\Lambda=0.7$ 
and $H_0=71$ km s$^{-1}$ Mpc$^{-1}$\nocite{Spergel03}({Spergel} {et~al.} 2003). 
In Eq.(\ref{eq:dnds}), $z(S_\nu,L_\nu)$ is derived by inverting the relation $S_{\nu}=L_{\nu}
(1 + z)/4\pi D_L^2(z)$, where $D_L(z)$ is the luminosity distance.

\begin{figure}
\begin{center}
\leavevmode
\epsfxsize=8cm \epsfbox{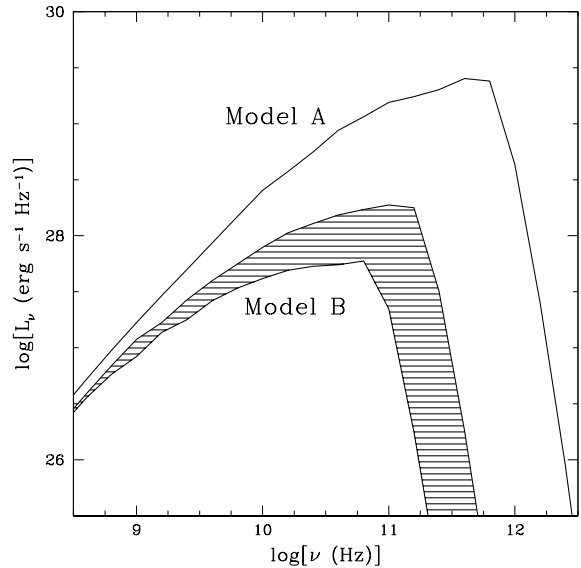}
\end{center}
\caption{ 
Synchrotron emission from low-radiative efficiency accretion flows.
The uppermost curve is the standard ADAF model with no outflows, while
the shaded region in between the two lower curves represent a range of ADAF models with large outflows.
The luminosity function in our model A is a log-gaussian with mean given
by that curve.  In the range of frequencies 30-100 GHz that we
are mostly interested in
 here, the average luminosity in the shaded region is $\sim 10^{28}$ erg/s/Hz. Our
model B is described by a log-gaussian with mean luminosity given by
this value.
}
\label{fig:emission}
\end{figure}

We will consider two models for the emission from accretion flows in 
early-type galaxies (see figure
\ref{fig:emission}): while these are computed on the basis of a
specific accretion model, they can be considered generally representative of the
typical range in luminosity and spectral shape that characterize these
sources. However, we should emphasize that, given the small sample of
observations so far, and the uncertainties in the specific model parameters
discussed below, our conclusions should be taken as indicative rather
than general and final.  Tighter constraints (which will leave less
space for parameter variations) will be obtained with future, more
sensitive CMB experiments together with a detailed, individual study
of a larger sample of sources.\\

a) Model A:

This is the standard ADAF model \nocite{ny}({Narayan} \& {Yi} 1994),
where the accretion rate is a constant function of the radius within
the flow.  In this model, the radio/microwave emission is due to
synchrotron emission from the inner regions of the accretion flow.
The emission at the self-absorbed synchrotron peak scales as $L_{\nu}
\propto \nu_{c}^2 T$, where $\nu_{c} \propto T^{2} B \propto T^2
\dot{M}^{1/2} M_{\rm BH}^{1/4} R^{-5/4}$ \nocite{ny2}({Narayan} \&
{Yi} 1995a); $T$ is the electron temperature, and $B$ the magnetic
field strength.  The ADAF model depends on a number of microphysical
parameters (i.e. the flow viscosity, the ratio of gas to magnetic
pressure, the adiabatic index of the fluid, the fraction of turbulent
energy which goes into heating the electrons) as well as geometrical
parameters (i.e. the radial extent of the flow, the possible
transition from a hot state to a cool disk at some radius), as well as
on the BH mass $M_{\rm BH}$ and on the accretion rate $\dot{M}$.  Whereas some constraints on
these parameters can be derived by multiwavelength observations and
broadband modeling of a given source, they cannot be well constrained
a priori and they can in principle vary from source to source.
Therefore, in a statistical study like ours, we can only adopt
``typical'' values for all these parameters.  In particular,
we use the ones adopted by \nocite{dimatteo}{Di
Matteo} {et~al.} (2000) for a BH of mass $\sim 10^9 M_\odot$ (as
typical of the bright end of the galaxy luminosity function that we
are considering here) and for an accretion rate $\sim 0.005$ in
Eddington units (note that the transition to an ADAF is believed (Narayan \& Yi 1995b) to
occur at rates below $\sim \alpha^2 \dot{M}_{\rm Edd}$, $\alpha$
being the viscosity parameter of the flow).
The luminosity predicted by this model is shown by the upper line in
Figure 1. To allow for a spread in BH masses, accretion rates,
 and parameters of the
flow around this typical value, we take the luminosity function
$f(L_\nu)$ to be a log-gaussian with mean given by the curve displayed
in the figure, and a standard deviation $\sigma$.  We adopt
$\sigma=0.25$; the predicted number counts are very little dependent
on the precise value of $\sigma$.\\

b) Model B:

The luminosity function for this model is calibrated on the sample of
those sources for which the broadband energy distribution was best fit
with an ADAF model with winds \nocite{dimatteo}({Di Matteo} {et~al.}
2000).  At frequencies in the range $\sim$ 30--100 GHz, the mean luminosity is 
roughly constant, and on the
order of $L_\nu\sim 10^{28}$ erg s$^{-1}$ Hz$^{-1}$ (see the dashed region in figure
\ref{fig:emission} which encompasses the range of those observations).
Hence, for this model
we assume the mean luminosity to be constant over the considered frequency range and
take the luminosity function $f(L_\nu)$ to be a log-gaussian with mean
given by the above value, and a standard deviation $\sigma=0.25$.

\begin{figure}
\begin{center}
\leavevmode
\epsfxsize=8cm \epsfbox{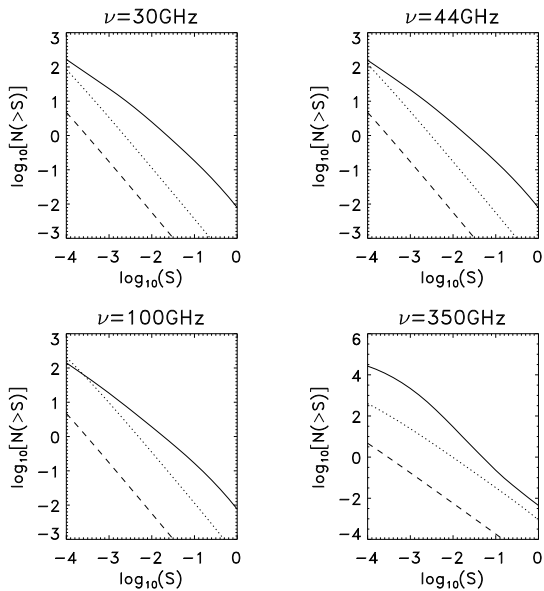}
\end{center}
\caption{ The number counts of accretion flow
 sources at different frequencies.
Long--dashed (short--dashed)  lines correspond to model A (B),
as described in the text.
The solid line are radio and infrared sources as reported from 
Toffolatti et al.~(1998), rescaled by a factor 0.8 (Argueso et al. 2003).
 Fluxes are in Jy, number counts in deg$^{-2}$.}
\label{fig:NgtS}
\end{figure}

Figure \ref{fig:NgtS} 
shows the number counts predicted by the sources in Model A
and B. 
 These sources are typically fainter than the
radio sources presented in \nocite{Toffo98}{Toffolatti} {et~al.} (1998),
and therefore the number counts are dominated by the local population.
As a consequence, 
the slope in figure \ref{fig:NgtS}
 is very close to Euclidean down to low flux levels.
Being faint, these sources 
are not  likely to be individually detected.  However, because of the
steepness of their number counts, they may contribute significantly to
the residual signal in  CMB experiments, where their relevance with
respect to the \nocite{Toffo98}{Toffolatti} {et~al.} (1998) type of sources becomes increasingly important 
as the detection flux limit of the survey is reduced.
   Moreover, because of the
radio spectrum of these sources is inverted, they are less likely to
be found at low frequencies, where most point sources catalogs are
compiled, and therefore   may
be neglected in present data analysis.
  In the following, we will estimate the contribution of these accretion
flow  sources to current CMB experiments, 
and explore the consequences that the emission from these (mostly
unresolved) point sources can have on the determination of the matter
power spectrum normalization $\sigma_8$ from these experiments.

\section{Point sources and CMB experiments}
Point sources are detected in CMB maps as high point-like
fluctuations above the mean.
Their detection depends on instrumental  properties such as the beam size and 
the  noise
level, as well as on the characteristics of the competing sky signals.
Assuming that all sources above a given flux limit $S_{lim}$ are subtracted
from the data, residual sources still contribute to the statistics of the map 
\nocite{Pierpa03}({Pierpaoli} 2003). 
For Poisson--distributed sources,
 the residual power spectrum and bispectrum are constant
for all scales and read:
\begin{equation}
\sigma^2 (\nu) = C_l(\nu) = g(x)^{2}~\int^{S_{lim}}_0~dS~ dn/dS~S^2
\end{equation}

\begin{equation}
\tilde{b}(\nu) = g(x)^3~\int^{S_{lim}}_0~dS~ dn/dS~S^3
\end{equation}
where 
 $g(x) = 2 {(hc)^2 /{(kT)^3}}~ 
\left[\sinh(x/2) / x^2\right]^2$  (with $(hc)^2 /{(kT)^3}=0.02 
\mu{\rm K  sr/Jy} $) and $x=\nu ({\rm GHz})/56.78$.

The bispectrum is sometimes quoted as the adimensional quantity 
$b\equiv\tilde{b}/T^3$.
If the sources are clustered, the power spectrum is increased by:
\begin{equation}
 C_l(\nu)_{clu} = w_l~(I(\nu))^2
\end{equation}
where the intensity $I(\nu) = g(x)~\int^{S_{lim}}_0~dS~ dn/dS~S$
and the $w_l$ are the Legendre expansion of the angular correlation
function:
\begin{equation}
 w(\theta) = w_l~P_l(\cos \theta).
\end{equation}
In order to compute the clustering term,
we use the recent determination of the SDSS red sample 
angular correlation function
\nocite{buda}({Budav{\' a}ri} {et~al.} 2003). This estimate 
should provide a conservative limit to the clustering 
contribution, since the red sample contains more galaxies than just
ellipticals, and ellipticals are more clustered than other sources.

Finally, given a particular experiment it is useful
to quote the  (non--clustered) power spectrum as
 the {\it noise per pixel} caused by the residual point sources as:
\begin{equation}
\sigma_{b}=(C_l/\theta_b^2)^{1/2}
\end{equation}
where $\theta$ is the pixel dimension, typically assumed as the
 FWHM of the beam.

In the following, we will estimate the contribution of these sources
to the current WMAP, CBI and BIMA signal.

The WMAP spectrum is clearly dominated by the primary CMB and,
given the  high flux cut ($ \simeq 1 $Jy),
the observed bispectrum  is likely to be dominated by other radio
sources more luminous than elliptical galaxies at these frequencies.
WMAP, however, presents the advantage of being an all--sky survey. 
Number counts of point sources which are confirmed to be elliptical
galaxies by mean of other (optical and infrared)
 observations may give a hint on the 
typical radio emission of early--type galaxies.

CBI and BIMA, on the contrary, observed a small area of sky with high
resolution and sensitivity.  These experiments detected the CMB power
spectrum at very high $l$'s, where the primary CMB is weak and the
point source contribution may be dominant. Therefore, the contribution
of the emission from accretion flows in early-type galaxies may be
significant to the observed CBI and BIMA power spectrum. We will
estimate the importance of this contribution and its consequences on
the derived cosmological parameter estimation.

\subsection{ WMAP detected sources}

The WMAP experiment \nocite{Bennet03}({Bennett} {et~al.} 2003)
 detected 208 point sources with a flux $S > 0.75$ in the V band.
 Among those sources, 29 have been identified as galaxies while 5 have
 no identification and could therefore be galaxies
 \nocite{Trushkin}({Trushkin} 2003).  Hence at most there is a total
 of 34 sources which are candidate for harboring a source of emission
 from an accretion flow.  This classification, however, does not
 distinguish elliptical galaxies from spirals.  In order to determine
 how many of the WMAP sources could be elliptical emitting according
 to the model described here, we proceeded as follows.  We selected the
 sources in \nocite{Trushkin}{Trushkin} (2003) according to the
 following two criteria: {\it i)} we eliminated sources with measured
 redshift greater than $z\sim0.01$, because if they were ellipticals
 emitting according to our models they would have an observed flux
 lower than 1 Jy; {\it ii)} we kept only sources with flat or inverted
 spectra.  We then visually inspected the remaining sources in the
 2MASS database, in order to select those which appear to be
 elliptical galaxies.  We only found two sources of this kind.  A
 comparison with the SDSS data (which currently covers about 10 per
 cent of the sky) finds only one elliptical galaxy among the WMAP
 sources.  A total number of 2--3 ellipticals in the WMAP sample of
 detected sources seems a conservative estimate.

 Model A predicts about 4.5 (8) sources above a flux $S_{lim}=1
 (0.75)$ Jy at 44 GHz, while model B would predict 0.15 (0.25) sources
 for the same flux cut.  These predictions seem to favour model A,
 model B predictions being 4--6 $\sigma$ away from observations.
 Current data therefore appears to suggest that the bulk of the
 elliptical galaxy population emits at $\sim$ 40 GHz more than the
 $\sim 10^{28}$ erg s$^{-1}$ Hz$^{-1}$ level assumed in model B.

Planck should reduce the source detection threshold of at least one order of magnitude
with respect to WMAP, possibly allowing for the detections of 
hundreds of these sources.
Moreover,  SDSS at completion should observe about one quarter of the sky.
By comparing these two data sets it will be possible to infer more
conclusive statements on accretion 
models purely by means of radio number counts.

\subsection{Current limits from CBI and BIMA}
Recent results on the  CMB power spectrum from
 high resolution CMB experiments  around 30 GHz like CBI and BIMA 
have shown a rise in the
high $l$'s  which is interpreted as  signature of the 
 SZ power spectrum from galaxy clusters \nocite{KS02,Bond02}({Komatsu} \& {Seljak} 2002; {Bond} 2002).
This interpretation allows to  infer a  
  value of the matter power spectrum $\sigma_8 \simeq 1$ \nocite{Read04}({Readhead} 2004).
Residual radio point sources, however, are also likely to contribute 
to the observed power spectrum.
We argue here that accretion flow sources may present a sizable 
contribution to the CBI and BIMA power spectra.

In CBI and BIMA individual point sources have been subtracted from the data
when identified by means of other observations of 
the same sky area at much lower 
frequencies.

In the CBI case, point sources 
above the flux limit of 3.4 mJy in the NRAO VLA Sky Survey (NVSS) at 1.4 GHz 
have been subtracted from the data.
A residual point source signal, compatible with the observed
point source population above the flux cut, has been modeled in
the data analysis. 
The population of sources considered in our work, however, is not likely 
to be well represented by the detected one, because 
our sources are typically very faint at such low frequencies.
As a result, the observed population may be rather biased toward
flat or falling spectra sources.
The observed population, in fact, shows a  less steep $dn/ds$ ($\propto 
S^{-1.875}$)
than the one appropriate for the point sources considered here ( $\propto S^{-2.5}$), and a mean 
spectral index   $\alpha \simeq -0.45$, 
while our sources have $\alpha = 0.5-0.8$
in the relevant frequency range of $\sim$ 1-10 GHz.
We extrapolate the NVSS flux limit to 30 GHz using a conservative
spectral index  $\alpha = 0.5$, and obtain for CBI $S_{\rm lim,30 GHz}=16$ mJy
as a nominal flux limit for our sources. 

In the case of BIMA, sources are detected with the VLA 
 at 4.8 GHz  with a flux limit of 150-175 $\mu$Jy \nocite{Dawson02}({Dawson} {et~al.} 2002).
If radio sources  have  a flat or falling spectrum, then this procedure
ensures the subtraction of sources at 30 GHz down to a very low flux limit.
For this reason, the BIMA data analysis 
did not  consider a  possible contribution from residual point sources.

Because accretion flow  sources present an inverted spectrum, we do expect 
a residual signal from them below the flux limit at 30 GHz. 
Considering the conservative 
case of a spectral index of 0.5, such limit is $S_{lim,30{\rm GHz}}=0.4 {\rm mJy} $

\begin{figure}
\begin{center}
\leavevmode
\epsfxsize=8cm \epsfbox{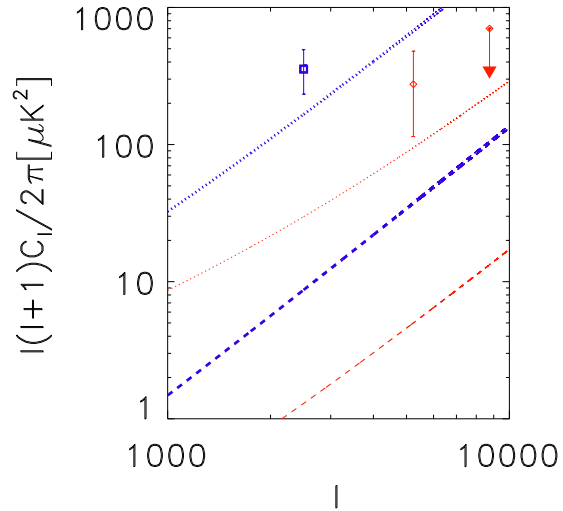}
\end{center}
\caption{ Residual power spectrum 
at 30 GHz from accretion flow sources and the high--l points from
the CBI  and BIMA  experiments. Bold--blue (thin--red)
 lines are relative to the
CBI (BIMA) flux cut; while the square--blue (triangle--red)
 points are the CBI (BIMA) results
with 1 $\sigma$ error bars.
The $l \simeq 9000$ BIMA point is a 90 per cent upper limit.
 Dotted (dashed) lines refer to model A (B)
respectively, as in figure 1. }
\label{fig:CBIBIMA}
\end{figure}

Given the considerations above, we assumed that all sources above the
 mentioned flux limits at 30 GHz have been subtracted from CBI and
 BIMA data, but the residual signal has not been taken into account.
 We computed the power spectrum 
from residual point sources given the
 above mentioned flux limits (relative to the two experiments) and the
 two different models. We also took into account 
the sources correlation which contribute
 about 15--20 per cent. We then compared the residual power spectra
 with the CBI band power for $l > 2000$ and the BIMA ones at $l >
 5000$.  Our results are presented in fig.~\ref{fig:CBIBIMA}.  We find
 that, in the case of the emission at the level of model B for all
 sources, their contribution would be 2.4\% for CBI, and 2.5\% for
 BIMA while, in the case of model A the contribution would be 47\%
 to CBI and 46\% to BIMA. Since $C_l^{SZ} \propto \sigma_8^7$,
  in the latter case the unresolved signal
 from these sources could produce a bias of $6-7\%$ in the current
 determination of $\sigma_8$ with this method \nocite{KS02,Bond02}({Komatsu} \& {Seljak} 2002).  The real value
 of $\sigma_8$ is still a very controversial issue, but the one
 inferred from SZ power spectrum is certainly in the high end of the
 considered range.  A lower $\sigma_8$ value would be in better
 agreement with determinations from cluster abundance
 \nocite{PBSW}({Pierpaoli} {et~al.} 2003), the WMAP results
 \nocite{Spergel03}({Spergel} {et~al.} 2003) and some weak lensing
 experiments \nocite{Heymans04}({Heymans} {et~al.} 2004).

\begin{table}
\begin{center}
\begin{tabular}{ccccc}
$\nu$ (GHz) & $S_{lim}$ (Jy) & $b_{A}$ & $b_{B} $ & $b_{\rm T}$ \\ \hline
30 & 1 & 7.0e-25 & 3.9e-26 &6.4e-23 \\
30 & 0.01 &6.9e-28 & 3.9e-29 &1.4e-26  \\
30 & 0.001 &1.9e-29 & 1.2e-30 & NA  \\
44 & 1 &  1.2e-25  & 4.3e-27 &7.2e-24 \\
44 & 0.01 & 1.1e-28 & 4.3e-30 & 1.4e-27\\
44 & 0.001 & 3.1e-30 & 1.3e-31 & NA\\
100 & 1 & 3.2e-27   & 5.7e-29 &9e-26\\
100 & 0.01 &  3.0e-30 & 5.7e-32& 1.8e-29 \\
100 & 0.001 &  7.5e-32 & 1.7e-33& NA \\
350 & 1 & 4.5e-27 & 2.8e-29 &7e-26\\
350 & 0.01 & 3.9e-30 & 2.8e-32 & 4.e-28\\ 
350 & 0.001 & 8.3e-32 & 8.4e-34 & NA\\ \hline
\end{tabular}
\end{center}
\caption{ Bispectrum  values implied by residual accretion-flow sources.
Column 1 is the frequency of observation, column 2 is the assumed flux limit,
 columns 3 and 4 are bispectrum 
estimates  for point source models A and B; 
column 4 are values reported from Argueso et al (2003) 
for  standard radio/IR sources (Toffolatti et al. 1998).}
\label{tab:bisp}
\end{table}


\subsection{Residual signal in all--sky experiments}
In this section we discuss the residual contribution from accretion flow
sources to the 
 bispectrum of WMAP and Planck.

In table~\ref{tab:bisp} we report the bispectrum produced by residual
 sources for different flux cuts and normalizations, compared with the
 one implied by more standard radio and infrared (IR) population as in
 the model by \nocite{Toffo98}{Toffolatti} {et~al.} (1998) (see also
 \nocite{Argueso03}{Arg{\" u}eso} {et~al.} (2003)) \footnote{It should
 be kept in mind that these quoted values have been reduced by a
 factor 0.8 in order to match the WMAP observations (Argueso et
 al. 2003).}.  WMAP measured the bispectrum in Q (40 GHz) and V (61
 GHz) bands, and found $\tilde{b}= (9.5 \pm 4.4)\times 10^{-5} \mu{\rm
 K^3 sr^2}$ and $\tilde{b}= (1.1 \pm 1.6)\times 10^{-5} \mu{\rm K^3
 sr^2}$ respectively.  For a nominal flux cut at $S_{lim}=0.75$
 accretion flow sources would imply $\tilde{b}= 2.9\times 10^{ -7}
 (1.1\times 10^{ -8}) \mu{\rm K^3 sr^2} $ in the V band for model A
 (B) and $\tilde{b}=1.7\times 10^{ -6} (6.0\times 10^{ -8}) \mu{\rm
 K^3 sr^2} $ in the Q band, so they contribute at most at the ***25
 per cent level in the V band, and 2 per cent in the Q band.  Given
 the WMAP bispectrum errors bars, these numbers do not allows us to
 draw constraints on accretion flow sources from these data.  The
 current flux limit from WMAP is so high ($S_{lim} \simeq 1$~Jy) that
 the contribution from these sources is sub dominant with respect to
 what is inferred by more standard radio/IR sources.  The observed
 bispectrum in the current WMAP data is consistent with a residual
 signal produced by the population of point sources observed at fluxes
 $S > 1$~Jy \nocite{komatsu03}({Komatsu} {et~al.} 2003), which is not
 dominated by accretion flow sources, rather by quasars and active
 galaxies as in the \nocite{Toffo98}{Toffolatti} {et~al.} (1998) model
 (see fig.~\ref{fig:NgtS}).

Because of the high sensitivity of Planck, point sources are expected
to be detected down to a flux between 0.01 and 0.1 Jy.  In
table~\ref{tab:bisp} we show the contribution of accretion flow
sources to the Planck bispectrum.  Their contribution at 100 GHz can
be $\sim 15-20\%$ the one from the \nocite{Toffo98}{Toffolatti}
{et~al.} (1998) radio/IR point sources population as computed in
\nocite{Argueso03}{Arg{\" u}eso} {et~al.} (2003).  In addition, Planck
will observe over a broad frequency range, allowing a better
description of point sources properties and possible discrimination
between accretion-flow sources and other sources.

\subsection{Future high--resolution SZ surveys}
Future high resolution SZ surveys in the 150--350 frequency range
 like ACT or APEX will allow
to detect point sources down to a flux limit $S_{lim} \simeq 1$ mJy.  
Because this point source population presents an emission that extends 
 to  very high frequencies (see fig. \ref{fig:emission}),
it may give a significant contribution to the observed signal in 
future SZ surveys.
Here we report the estimated pixel noise produced 
by residual sources for different flux 
cut and for a fiducial pixel of $1^{\prime}$.
Again we report our results for the two models shown in Figure 1.
These two figures should roughly provide a lower and upper limit to the 
contribution coming from elliptical galaxies.
 Our results are summarized in table \ref{tab:noise} together
with the estimates of \nocite{WM03}{White} \& {Majumdar} (2004)  
for the population of currently observed radio sources.
For the sake of comparison, here we neglected the contribution
of clustering.
Accretion flow sources may produce a noise that is $\sim 25-30$\%
of the one produced by the population of observed sources.

At the same frequencies, infrared sources are currently
 predicted to produce a slightly higher signal than radio sources \nocite{WM03}({White} \& {Majumdar} 2004).
These IR estimates, however,  may be biased high if part of the 
signal is due to lensing effect.
Future experiments like SCUBA--2 and the Herschel Space Observatory 
will help in better  characterizing the IR population. 

\begin{table}
\begin{center}
\begin{tabular}{ccccc}
$\nu$ (GHz) & $S_{lim} (mJy)$ & $\sigma_{b,A}$ & $\sigma_{b,B}$ & $\sigma_{b,r}$ \\
\hline \\
      150  &     1 &      1.3   &  0.3  &  5 (4)\\
      150  &    5  &     2.2  & 0.5  &  9 (7)\\
      150  &    10 &      2.8  & 0.6  &  12 (9)\\
      150  &    50 &      2.9  & 0.9  &  20 (16)\\
      220  &    1 &      1.1  & 0.3  &  4 (3)\\
      220  &    5 &      2.0  & 0.4  &  7 (5)\\
      220  &    10 &      2.5  & 0.5  &  9 (7)\\
      220  &    50 &      2.6  & 0.7  &  16 (12)\\
\hline\\

\end{tabular}
\end{center}
\caption{Pixel noise implied by residual  accretion flow 
sources in SZ experiments with 
$1^{\prime}$ resolution. Columns 3 and 4 are predictions for model A and B.
 Column 5 are predictions from White and 
Majumdar (2003) on the standard radio sources contribution (the two numbers refer to 
their two different extrapolations from low frequencies). 
 The last three columns are
in $\mu$K for a nominal $1^{\prime}$ pixel.}
\label{tab:noise}
\end{table}

\section{Conclusions}

We investigated the possible contribution to current and future CMB
experiments from the accretion flows around supermassive black holes
believed to harbor the centers of galaxies.  These sources are
typically faint and numerous, therefore they are likely to be observed
as a residual signal in CMB experiments.  A comparison with
the number counts of WMAP detected sources (and identified as 
elliptical galaxies) indicates that the bulk of the emission
is probably higher that the $\sim 10^{28}$ erg/s/Hz assumed
in our emission model B.

 We have showed that
 the residual signal produced by the emission from these unresolved sources could
significantly contribute to the CBI and BIMA observed power spectrum.
More specifically, these sources could make up to $\sim 40-50\%$ of
the observed signal at $l > 2000$, therefore
reducing the inferred value of $\sigma_8^{SZ}$ by $6-7\%$.
As for all sky experiments, we showed that the residual signal in the
WMAP maps with the current flux limit is likely to be dominated by
other kinds of sources, with accretion-flow sources contributing at
the 2-3 per cent level at most.

Planck will have a better sensitivity than WMAP and will allow to
detect fainter sources reducing the flux limit by at least a factor of 10
and possibly detecting hundreds of early--type galaxies.
We showed that the residual signal from accretion-flow sources may
contribute up to the 15\% to the bispectrum around 100 GHz.
Unlike other radio sources, these sources typically
show an inverted spectrum and are quite bright also in the infrared,
up to about 300 GHz.  This feature may facilitate their
detection and discrimination with respect to the standard radio and
infrared sources, especially with Planck that offers the potential of
an all--sky experiment with a broad frequency coverage.
A characterization of the Plank sources with the SDSS will allow
to clearly distinguish between elliptical galaxies and other kinds
of sources, allowing to put constraints on accretion flow models with
radio number counts.

Future high--resolution CMB experiments like ACT and APEX operating in
the 150--350 GHz range may also be affected by the emission from
accretion flows in galaxies.
The residual power spectrum is predicted to be about 25-30\% of
the one expected for currently observed radio point sources.
 At these
frequencies, however, IR sources are likely to be dominant, unless
their number count is biased high due to lensing effects.  Future
surveys with  instruments like SCUBA--2 and Herschel
will help in clarifying this issue.

\section*{ACKNOWLEDGMENTS}

EP is an NSF-ADVANCE fellow (NSF-0340648) also supported by 
NASA grant NAG5-11489.
The authors wish to thank Luigi Toffolatti for providing his data on
number counts; Colin Borys, Francisco Argueso and Tiziana Di Matteo, Idit
Zehavi and Tamas Budavari, Zeljko Ivezic for useful   conversations; 
Luigi Danese and Gianfranco De Zotti for useful feedback on the manuscript.



\end{document}